\documentclass[12pt]{article}
\usepackage{a4wide}
\usepackage{amssymb}
\usepackage{amsmath}
\usepackage{graphicx}
\usepackage{mathdots}
\usepackage{slashed}
\usepackage{verbatim}
\usepackage{bm}
\usepackage{cite}
\usepackage{xcolor}
\usepackage[utf8]{inputenc}
\usepackage{mathtools}

\usepackage[normalem]{ulem}

\usepackage{hyperref}
\usepackage{cleveref}

\usepackage{IEEEtrantools}

\def\makeatletter{\catcode`\@=11}
\makeatletter
\def\mathbox#1{\hbox{$\m@th#1$}}%
\def\math@ccstyles#1#2#3#4#5#6#7{{\leavevmode
      \setbox0\mathbox{#6#7}%
      \setbox2\mathbox{#4#5}%
      \dimen@ #3%
      \baselineskip\z@\lineskiplimit#1\lineskip\z@
      \vbox{\ialign{##\crcr
             \hfil \kern #2\box2 \hfil\crcr
             \noalign{\kern\dimen@}%
             \hfil\box0\hfil\crcr}}}}
\def\mathaccstyles{\math@ccstyles\maxdimen}
\def\maththroughstyles{\math@ccstyles{-\maxdimen}}
\def\unity%
 {\maththroughstyles{.45\ht0}\z@\displaystyle {\mathchar"006C}\displaystyle 1}
 
\usepackage{tikz}
\usetikzlibrary{positioning,arrows}
\usetikzlibrary{decorations.pathmorphing,calc}
\usetikzlibrary{decorations.markings}
\newif\ifmirrorsemicircle
\usepackage{float}
\usepackage{ifthen}
\usepackage{enumerate}
\usepackage{amsmath}
\usepackage{amssymb}
\usepackage[mathscr]{eucal}
\usepackage[errorshow]{tracefnt}
\usepackage{amsmath}
\usepackage{slashed}
\usepackage{amssymb}
\usepackage{array}
\usepackage{amsthm}
\usepackage{eucal}
\usepackage{epsf}
\usepackage{epsfig}
\usepackage{psfrag}
\usepackage{multirow}
\usepackage{wasysym} 
\usepackage{tikz-cd} 
\usepgfmodule{shapes}
\usetikzlibrary{backgrounds, arrows,calc,shapes,decorations.pathreplacing, decorations.markings, automata,positioning}
\usepackage{verbatim}

 \numberwithin{equation}{section}
\begin{document}

\begin{flushright}\footnotesize

\texttt{}
\vspace{0.6cm}
\end{flushright}

\mbox{}
\vspace{0truecm}
\linespread{1.1}

\centerline{\LARGE \bf Non-Abelian R-symmetry and dielectric branes}
\medskip

\vspace{.5cm}

 \centerline{\LARGE \bf }

\vspace{1.5truecm}

\centerline{
    { \bf Francesco Mignosa${}^{a,b}$} \footnote{francesco.mignosa02@gmail.com}
        {\bf and}
    { \bf Diego Rodriguez-Gomez${}^{a,b}$} \footnote{d.rodriguez.gomez@uniovi.es}}

\vspace{1cm}
\centerline{{\it ${}^a$ Department of Physics, Universidad de Oviedo}} \centerline{{\it C/ Federico Garc\'ia Lorca  18, 33007  Oviedo, Spain}}
\medskip
\centerline{{\it ${}^b$  Instituto Universitario de Ciencias y Tecnolog\'ias Espaciales de Asturias (ICTEA)}}\centerline{{\it C/~de la Independencia 13, 33004 Oviedo, Spain.}}
\vspace{1cm}

\centerline{\bf ABSTRACT}
\medskip 
 
We study the holographic realization of the $SO(6)_R$ R-symmetry of $\mathcal{N}=4$ super Yang-Mills with unitary gauge group. Focusing on 1/2 BPS states in the $[0,J,0]$ representation of $SO(6)_R$, it is known that depending on the scaling of $J$ with $N$, these are best described holographically in terms of gravitational waves ($J\ll N$) or D3 brane giant gravitons ($J\sim N$). These two descriptions are bridged by the dielectric effect, as the D3 giant can be regarded as a puffed-up configuration of gravitational waves. The natural non-BPS branes for symmetry operators are either 4-branes or non-BPS Kaluza-Klein monopoles. We show that the former can be regarded as a dielectric expansion of the latter, in parallel to the charged operators. We also propose symTh and symTFT candidates for the $SO(6)_R$ symmetry, whose operators at the boundary must correspond to the non-BPS branes.

\noindent

\newpage

\tableofcontents

\section{Introduction} \label{intro}

Symmetries are fundamental to investigating the dynamics of quantum field theories (QFTs). They allow to classify the spectrum of the QFT, as well as constrain the physical processes by imposing selection rules. When they suffer from 't Hooft anomalies, they can give insight into the RG flow of the corresponding QFT through 't Hooft anomaly matching. Moreover, in the continuous case, their spontaneous breaking predicts the existence of massless particles through Goldstone's theorem. Over the last decade, it has become clear that a particularly useful approach to symmetries is by associating them with a sector of topological defects that act on the charged operators through their linking. Based on this new paradigm, the basic notion of symmetry has been extended to a higher-form case, whose symmetry operators are supported on submanifolds of arbitrary codimension. Moreover, the topological symmetry operators do not necessarily fuse according to a group law, belonging more generally to a higher-dimensional category. 

Along an intimately related line, it has become clear the usefulness of separating the symmetries from the dynamics of an associated QFT by encoding the former in a symmetry Topological Field Theory (symTFT) \cite{Freed:2012bs, Gaiotto:2020iye, Apruzzi:2021nmk, Freed:2022qnc}. For a QFT in $d$-dimensions, this consists of a $d+1$-dimensional topological auxiliary theory on a slab with two boundaries: a $d$-dimensional physical boundary, where the dynamics of the original QFT is localized, and a topological boundary, where different boundary conditions select different global forms of the symmetry of the original QFT.  Indeed, different boundary conditions allow different sets of topological operators of the symTFT to end at the topological boundary. While these are dual to the charged operators of the original QFT, operators which do not trivialize when brought to the topological boundary describe the topological operators of the original QFT. In the case of discrete symmetries, in particular with the aid of String Theory, the symTFT paradigm has been thoroughly studied. In contrast, the case of continuous symmetries is less established. For continuous Abelian global symmetries, the symTFT was constructed in \cite{Antinucci:2024zjp} (see also \cite{Brennan:2024fgj}). It has also been proposed an alternative description in terms of a non-topological symmetry Theory (symTh) \cite{Apruzzi:2024htg}. Very recently, the case of non-Abelian continuous symmetries was also considered in \cite{Bonetti:2024cjk,Jia:2025jmn,Apruzzi:2025hvs,Bonetti:2025dvm}

Specializing to holographic theories, it is natural to ask for the bulk realization of the symmetry operators of the QFT. A related, albeit independent question is the emergence of the symTFT/symTh from the bulk, which was explored in many instances, see \cite{OBtalks,Apruzzi:2022rei,GarciaEtxebarria:2022vzq,Bergman:2022otk,Heckman:2022muc,Apruzzi:2023uma, Bah:2023ymy, Heckman:2024oot, Argurio:2024oym, Cvetic:2024dzu,  Bergman:2024its, Heckman:2025lmw, Akhond:2025wwn, Bertolini:2025wyj}. In this last respect, at least for continuous symmetries, a mechanism to obtain the symTFT from holography has been recently proposed in \cite{Mignosa:2025cpg}.

Coming back to the bulk avatar of symmetry operators, in the case of continuous internal symmetries, it has been proposed that symmetry operators are realized through non-BPS branes in the bulk \cite{Bergman:2024aly}, which naturally relates to the fluxbrane construction of \cite{Cvetic:2023plv, Heckman:2025wqd}. This proposal has been thoroughly studied for baryon-like symmetries \cite{Calvo:2025kjh}, whose bulk gauge fields are encoded in RR potentials. The case of symmetries associated with isometries of the internal space has been recently discussed in \cite{Cvetic:2025kdn, Calvo:2025usj,Bah:2025vfu}. A distinctive feature of these is that they can be naturally non-Abelian. In \cite{Calvo:2025usj}, the simplest case of the $U(1)_R$ symmetry of holographic (toric) $\mathcal{N}=1$ 4d theories associated to the Reeb vector of the internal Sasaki-Einstein space was studied. It was proposed that the symmetry operators are realized through non-BPS Kaluza-Klein monopoles, which are 7d non-BPS objects living in a space with a transverse Killing direction and which decay into a standard KK monopole. This picture was shown to consistently reproduce the expected features of the R-symmetry, including its 't Hooft anomalies with the baryonic symmetry.

It remains a natural question to study the case of non-Abelian symmetries, notably the case of the $SO(6)_R$ R-symmetry of $\mathcal{N}=4$ SYM. Progress along these lines was made recently in \cite{Bah:2025vfu}. This note aims to add further ingredients of relevance to the description of this symmetry. More concretely, we show that, in addition to the non-BPS 4-branes discussed in \cite{Bah:2025vfu}, there are non-BPS KK monopoles as in \cite{Calvo:2025usj}. This results in some tension, as it appears that there are two different bulk constructions for operators realizing the same symmetry. As we show, this apparent redundancy is resolved by means of the dielectric effect. Indeed, remarkably, it turns out that the non-BPS KK monopole can be regarded as a dielectric expansion of 4-branes. This mimics the similar pattern for charged states found in \cite{Janssen:2003ri}.

The organization of this note is as follows. In section \ref{ChargedOps}, we study a natural set of operators in $\mathcal{N}=4$ SYM charged under the $SO(6)_R$ R-symmetry. These are scalars in the $[0,J,0]$ representation, whose highest weight state saturates the BPS bound $\Delta=J$. As it is well-known, depending on the range of $J$, the dual holographic description is either in terms of KK fluctuations of the geometry ($J\ll N$) or in terms of giant/dual giant gravitons ($J\sim N$), which are D3 branes spinning in AdS$_5\times S^5$. The KK fluctuations correspond to fluctuations of the geometry whose associated particles are probe gravitational waves (GW). It is well-known that multiple coincident GWs undergo a dielectric expansion into D3 giant gravitons, thus bridging the two regimes. In section \ref{SymmetryOps} we turn to the symmetry operators. Elaborating on \cite{Bah:2025vfu}, we propose that the symmetry operators include the non-BPS KK monopoles of \cite{Calvo:2025usj} as well as 4-branes \cite{Bah:2025vfu}. In section \ref{dielectric} we argue that, analogously to charged states, coincident 4-branes undergo a dielectric expansion into non-BPS KK monopoles. In section \ref{symTFT}, we propose the symTh and symTFT for the $SO(6)_R$ R-symmetry, which should be behind the charged and symmetry operators that we found. We conclude in section \ref{conc} with some final comments. We compile some relevant technical details of the $S^5$ metric in Appendix \ref{symTh}.

\section{Charged operators}\label{ChargedOps}

We are interested in the $SO(6)_R$ R-symmetry of $4d\,\mathcal{N}=4$ SYM with unitary gauge group of rank $N$. The Cartan torus of $SO(6)_R$ is $U(1)^3$. The theory is dual to type IIB on AdS$_5\times S^5$ with an additional $F_5$ flux. 
Geometrically, each such $U(1)$ acts rotating three orthogonal planes (say $12$, $34$ and $56$) in the $\mathbb{R}^6$ embedding space of the $S^5$ parametrized by the coordinates $\mu^I, \, I=1,...,6$. When regarded as a $\mathcal{N}=1$ theory, the combination of the associated conserved currents that sit in the stress-energy tensor multiplet is $\frac{2}{3}$ of the sum of the three conserved currents related to the three orthogonal rotations.

A very natural set of operators charged under the R-symmetry symmetry are $\frac{1}{2}$ BPS operators $O_J$ made out of the scalars inside the three chiral superfields (in $\mathcal{N}=1$ language) of the theory. These come in representations of $SO(6)$ with Dynkin labels $[0,J,0]$. The highest weight state has charge $J$ under a $U(1)$ inside the Cartan torus, and saturates the BPS bound equating dimension and R-charge $\Delta=J$. Depending on the scaling of $J$ with $N$, the operators are differently realized in holography, as shown in Table \ref{JNscaling}.

\begin{table}[h!]
\centering
\begin{tabular}{|c|c|}
\hline
$J\ll \sqrt{N}$ & Supergravity KK fluctuation \\ \hline
$J\sim \sqrt{N}$ & Macroscopic string spinning \\ \hline
$J\sim N$ & Giant/dual giant graviton \\ \hline
$ J \sim N^2$ & Black hole in $AdS$ \\ \hline
\end{tabular}
\caption{Holographic description of $\frac{1}{2}$ BPS operators depending on their dimension/charge}
\label{JNscaling}
\end{table}

In particular, the dimension of the operator is also related to its mass as $\Delta=mR$  with $R$ the radius of the $S^5$ sphere. 
In the following, we will restrict to operators not so heavy as to change the background geometry, which is simply AdS$_5\times S^5$. The unit $S^5$ can be described by embedding coordinates $\mu^I$ into $\mathbb{R}^6$ subject to the constraint $\mu^I\mu^I=1$. The metric is then $d\mu^Id\mu^I$, and explicitly shows the $SO(6)_R$ as an isometry. For later purposes, it will be relevant to consider the gauging of the isometries. This is done standardly (see \textit{e.g.} \cite{Cvetic:2000nc}) by promoting 
\begin{equation}
d\mu^I\rightarrow D\mu^I=d\mu^I+A^{IJ}\mu^J\,,
\end{equation}
with $A^{IJ}=-A^{JI}$ the $SO(6)_R$ isometry gauge fields (see appendix \ref{symTh}).\footnote{Note that here we are reabsorbing the coupling $g=R_{S^5}^{-1}= (4\pi g_s N)^{-1/4} \sqrt{\alpha'}$ in the definition of the R-symmetry gauge fields.} 

 In the following two particular parametrizations of the $S^5$ will be useful. One of them is
\begin{equation}
\label{alphacoords}
 \mu^1+i\mu^2=\cos\theta\,e^{i\phi}\,,\qquad  \mu^3+i\mu^4=\sin\theta\,\cos\frac{\alpha_1}{2}\,e^{i\frac{\alpha_3+\alpha_2}{2}}\,,\qquad  \mu^5+i\mu^6=\sin\theta\,\sin\frac{\alpha_1}{2} \,e^{i\frac{\alpha_3-\alpha_2}{2}}\,,
 \end{equation}
 with $\theta\in \left[0,\frac{\pi}{2}\right]$,  $\phi\in\left[0,2\pi\right]$,  $\alpha_1\in\left[0,\pi\right]$,  $\alpha_2\in\left[0,2\pi\right]$ and  $\alpha_3\in\left[0,4\pi\right]$. In these coordinates the angles corresponding to the $U(1)^3$ Cartan torus of the $SO(6)_R$ rotating the  three planes are essentially $\{\phi,\alpha_2,\alpha_3\}$. In particular, the angle $\phi$ rotating the $12$ plane is directly related to a $U(1)$ inside the Cartan torus whose charge is $J$ above. Finally, in these coordinates, the metric of the 5-sphere is
\begin{equation}
\label{metricalpha}
ds_{S^5}^2=d\theta^2+\cos^2\theta d\phi^2+\sin^2\theta\,ds_{S^3}^2\,,\,\,\,\,\,\, ds_{S^3}^2=\frac{1}{4}\big[ d\alpha_1^2+\sin^2\alpha_1\,d\alpha_2^2+(d\alpha_3+\cos\alpha_1d\alpha_2)^2\big]\,.
\end{equation}
The other parametrization is
\begin{equation}
\label{Reebcoords}
\mu_1+i\mu^2=\cos\varphi_1 e^{i\chi}\,,\qquad \mu^3+i\mu^4= \sin\varphi_1 \cos\frac{\varphi_2}{2}e^{i \frac{2\chi+\psi+\varphi_3}{2}}\,,\qquad \mu^5+i\mu^6=\sin \varphi_1\sin\frac{\varphi_2}{2} e^{i \frac{2\chi+\psi-\varphi_3}{2}}\,,
\end{equation}
with $\chi\in [0,2\pi), \varphi_1\in [0,\frac{\pi}{2}], \varphi_2\in [0,\pi], \varphi_3\in [0,2\pi], \psi\in [0,4\pi)$. In these coordinates, the $S^5$ metric is
\begin{equation}
\label{metricfibration}
ds^2_{S^5}= (d\chi+\xi)^2+ ds^2_{\mathbb{C}P^2} 
\end{equation}
with
\begin{align}
& ds^2_{\mathbb{C}P^2}=d\varphi_1^2+ \frac{\sin^2\varphi_1}{4} \left[\cos^2\varphi_1(d\psi +\cos\varphi_2d\varphi_3)^2+d\varphi_2^2+\sin^2\varphi_2 d\varphi_3^2 \right] \,,\\
& \xi=\frac{1}{2}\sin^2\varphi_1(d\psi+\cos\varphi_2d\varphi_3)\,.
\end{align}
Note that the K\"ahler form $\mathcal{J}$ of $\mathbb{C}P^2$ is related to $\xi$ as  $2\mathcal{J}=d\xi$. These coordinates put the $S^5$ on equal footing with generic Sasaki-Einstein manifolds $X_5$, which are generically a $U(1)$ bundle over a four-dimensional K\"ahler-Einstein manifold $B$. These manifolds can be used to construct AdS$_5\times X_5$ solutions to Type IIB supergravity dual to four-dimensional superconformal QFT's generically with $\mathcal{N}=1$ supersymmetry. The metric of $X_5$ can be written as
\begin{equation}
\label{X5metric}
ds_{X_5}^2=g_{\chi\chi}\,(d\chi+\xi)^2+ds_B^2\,,
\end{equation}
into which eq. \eqref{metricfibration} immediately fits. In these coordinates, the isometry associated with the angle $\chi$ parametrizing the $U(1)$ fiber\footnote{We assume a normalization where $\chi$ is $2\pi$-periodic.} corresponds to the $\mathcal{N}=1$ R-symmetry sitting in the multiplet of the stress-energy tensor \cite{Berenstein:2002ke}. In $\mathcal{N}=1$ language, the charge $R$ of chiral primaries under this R-symmetry satisfies the BPS bound $\Delta=\frac{3}{2}R$.

For future reference, note that the coordinates in eq. \eqref{alphacoords} are related to eq. \eqref{Reebcoords} as
\begin{align}
\varphi_1= \theta\,,\qquad  \varphi_2=\alpha_1\,, \qquad \varphi_3=\alpha_2\,,\qquad \chi=\phi\,,\qquad  \psi= \alpha_2+\alpha_3-2\phi\,.
\end{align}

\subsection{Small dimension: $J\ll N$}

Operators with $J\ll \sqrt{N}$ are holographically realized as KK fluctuations of the dual supergravity background. The would-be particle associated with them is a probe gravitational wave (GW) moving in the internal geometry as dictated by the charges of the operator. The worldvolume action for a type IIB GW with momentum $m$ was constructed in \cite{Janssen:2002vb}, and reads
\begin{align}
\label{GWIIB}
& S_{\rm GW}= -m \, \int_{M_1} d\tau \kappa^{-1}\sqrt{|\partial_\tau X^M \partial_\tau X^N \mathcal{G}_{MN}|}+ m\, \int_{M_1} \imath_{\kappa}\tilde{B}_2\,, \\
& \mathcal{G}_{MN}=g_{MN}-\frac{\kappa^M\,\kappa^N}{\kappa^2}\,,
\end{align}
where $\kappa$ is a Killing vector pointing along the direction of propagation of the wave and $\tilde{B}_2$ is the NSNS 2-form associated with the background obtained from T-duality along $\kappa$ of the type IIB solution.\footnote{As we will often need to consider T-dual frames, we will denote by $\tilde{\cdot}$ quantities in the T-dual background.}

Following \cite{Calvo:2025usj}, we can make the correspondence between probe GW's and KK fluctuations rather explicit even for general $X_5$. Starting with eq. \eqref{X5metric}, we can gauge the isometry shifting $\chi$ to include the ($\mathcal{N}=1$) $U(1)$ R-symmetry gauge field (see eq. \eqref{gperturbedReeb}). In this background we consider a GW with $\kappa^{M}=\delta^{M}_\chi$, so that
\begin{equation}
\label{NSNSfielddual}
\tilde{B}_2= -\left(\xi+ \frac{2}{3}g_{\chi\chi}^{-1/2}A\right)\wedge d\chi\,.
\end{equation}
From the DBI part of the action in eq. \eqref{GWIIB} we can read off the mass of the particle, which corresponds to its dimension, as $\Delta=m/\sqrt{g_{\chi\chi}}$. In turn, the WZ part of the action is

\begin{equation}
\label{SGWReeb}
S_{\rm GW}\supset -\frac{2}{3}\,m\,g_{\chi\chi}^{-1/2}\,\int_{M_1} A\,.
\end{equation}
This shows that the GW is a Wilson line of $A$ with charge $R=\frac{2}{3}\,m/\sqrt{g_{\chi\chi}}$, which is precisely the $\mathcal{N}=1$ BPS bound. 

Specializing to the case of the $S^5$, we can alternatively consider, in the coordinates of eq. \eqref{alphacoords}, a GW propagating along $\phi$ with $\kappa^M=\delta^M_{\phi}$. As reviewed in the Appendix, we can easily turn on the $SO(6)_R$ gauge field fluctuations. Using eq. \eqref{gperturbed}, in this case

\begin{equation}\label{NSNSfielddualS5}
\tilde{B}_2= \frac{1}{2}K_{IJ}^{\phi}A^{IJ}\wedge d\phi\,.
\end{equation}
 Hence, it follows that the GW action contains

\begin{equation}
\label{SGWphi}
S_{\text{GW}}\supset \frac{1}{2}m\,\int_{M_1}\,K_{IJ}^{\phi}A^{IJ}\,.
\end{equation}
While the explicit expression for $K_{IJ}^{\phi}$ in eq. \eqref{Kphi} is rather non-illuminating; it is important that it contains contributions from $A^{1I}$ and $A^{2I}$, reflecting that the GW spins in the 12 plane and reacts to moving along transverse coordinates. Assuming that only $A^{12}$ is excited, from eq. \eqref{Kphi} we have $K_{IJ}^{\phi}A^{IJ}/2=A^{12}$, which is the counterpart of eq. \eqref{SGWReeb} for the case of the charge $J=m\ll N$.

It is interesting to review how the worldvolume action for GW in \cite{Janssen:2002vb} was constructed. The starting point is a rigid string wound $m$ times around an isometric direction. Upon T-duality, one finds the action for a particle, due to the rigidity of the string,  with $m$ units of momentum along the isometric direction (in fact, this makes apparent the appearance of $\imath_{\kappa}\tilde{B}_2$ in eq. \eqref{GWIIB}). This suggests that, at least to leading order, GW can cover the intermediate regime of $J\sim \sqrt{N}$, where the appropriate holographic description is really through macroscopic strings spinning in the holographic background, which, to leading order, are rigid.

\subsection{Large dimension: $J\sim N$}

For $J\sim N$, the operator is best described holographically as a giant/dual giant graviton. This is a D3 brane spinning either in the $S^5$ (giant) or the AdS$_5$ (dual giant). In the following, we will focus on giant gravitons, as the description of these does not require committing to the choice of global coordinates for AdS$_5$. Giants are best described in the coordinates reported in eq. \eqref{alphacoords}. Considering a D3 brane wrapping the $S^3$ in eq. \eqref{metricalpha} and spinning along $\phi$ in the plane $12$ with momentum $J$ at constant $\theta$, one easily sees that its mass/dimension of the dual operator is $\Delta=J$, with $J=N\sin^2\theta$ \cite{McGreevy:2000cw,Grisaru:2000zn,Hashimoto:2000zp}. Note that $J$ is naturally of order $N$.

Just as before, we can make the correspondence between the particle and the corresponding KK fluctuation precise. Upon turning on the $SO(6)_R$ gauge fields as fluctuations, it is straightforward to compute the action for the giant graviton, finding, to linear order in the R-symmetry field
\begin{equation}
\label{D3action}
S_{\text{D3}}=-J  \,\int_{M_1}\,A^{12}\,.
\end{equation}
It is important to stress that the gauge fields for the $SO(6)_R$ not only appear in the metric but also in the RR 5-form field strength. In fact, the result in eq. \eqref{D3action} comes from a sum of two contributions, one coming from the DBI action and one from the WZ term of the D3 brane action. This is in sharp contrast with the case of GW's in the previous subsection, which only couple to metric fluctuations (the monopole coupling to $\imath_{\kappa}\tilde{B}_2$ is entirely a metric component of the original background).

Thus, \eqref{D3action} shows that giants can be regarded as Wilson lines for $A^{12}$, in the end, they couple to $A^{12}$ just like GW in eq. \eqref{SGWphi}, only that with $J\sim N$, and consequently, when ending on the boundary, correspond to operators with charge $J$ under a $U(1)$ inside the Cartan of the $SO(6)_R$ saturating the bound $\Delta=J$. 

Since in the end giant gravitons and GW describe essentially the same operator, only that with different dimension; it is natural to suspect that giant gravitons can be described as some aggregate of GW's. The precise relation between giants and GW's was made explicit in \cite{Janssen:2003ri}, where it was shown that $N$ coincident GWs expand through the dielectric effect \cite{Myers:1999ps} into giant gravitons. Crucially, to construct the action for multiple GW in Type IIB String Theory, one performs a T-duality of the Type IIA action for coincident GW along a transverse direction with Killing vector $\ell$ to that of propagation. As a consequence, the action depends on both $\kappa$ and $\ell$ as 
\begin{align}
& S_{\rm GW}=-T_0\,\int d\tau\,{\rm Str}\Big\{ \kappa^{-1}\,\sqrt{-P[E_{00}+E_{0i}(Q^{-1}-\delta)^i_k\,E^{kj}\,E_{j0}]\,{\rm det}(Q)}-i\,P[(\imath_X\imath_X)\,\imath_{\ell}C^{(4)}]\Big\}\,;\nonumber \\
& E_{MN}=g_{MN}-\frac{\kappa_M\kappa_N}{\kappa^2}-\frac{\ell_M\ell_N}{\ell^2}-\frac{e^{\phi}}{\kappa\ell}\,(\imath_{\kappa}\imath_{\ell}C^{(4)})_{MN}\,,\qquad  Q^M_N=\delta^M_N+i\,[X^M,X^P]\,e^{-\phi}\kappa\ell\,E_{PN}\,.
\label{SdieletricGW}
\end{align}
Thus, in the coordinates in eq. \eqref{alphacoords} one naturally identifies $\kappa$ with $\phi$. Moreover, $\ell$ is identified with $\alpha_3$, and plays the role of the isometry of the Hopf fibration of the $S^3$. Then, the dielectric effect comes from the expansion of the GWs into the $S^2$ base of the Hopf fibration. Parametrizing the Cartesian coordinates of this fuzzy $S^2$ through the generators  $J^i$ of an $N\times N$ representation of $SU(2)$ as
\begin{equation}
X^i=\frac{J^i}{\sqrt{N^2-1}}\,,\qquad [J^i,J^j]=2\,i\,\epsilon^{ijk}\,J^k\,,\qquad \vec{J}^2=(N^2-1)\,\unity\,,
\end{equation}
one can see that, in the large $N$ limit, the microscopic dielectric description exactly reproduces the macroscopic spinning D3 brane.

\section{Symmetry operators}\label{SymmetryOps}

Let us now turn to the symmetry operators for $SO(6)_R$. As these must link with the charged objects described above, we expect them to be represented by non-BPS branes wrapping, in AdS$_5$ a three-manifold $M_3$ linking with $M_1$, such that their worldvolume theory after the reduction contains a term proportional to $\int_{M_3}\star_{\text{AdS}_5} d\mathcal{A}$ for whatever $\mathcal{A}$ the charged object sources (be it $A^{12}$ for GW's/giants spinning along $\phi$ in the coordinates of eq. \eqref{alphacoords} or $A$, the $\mathcal{N}=1$ R-symmetry gauge field for  GW's/giants spinning along $\chi$ in the coordinates of eq. \eqref{Reebcoords}). Since the $SO(6)_R$ symmetry gauge fields appear in both the metric and the RR 5-form field strength, it is natural to identify as symmetry operators the non-BPS KK monopoles of \cite{Calvo:2025usj}, on top of the 4-branes described in \cite{Bah:2025vfu}.

\subsection{Non-BPS KK monopoles}

In \cite{ Calvo:2025usj} it was shown that non-BPS KK monopoles represent the symmetry operators for the $U(1)_R$ R-symmetry of holographic $\mathcal{N}=1$ superconformal QFT's. More recently, in \cite{Rodriguez-Gomez:2026mjj}, it was speculated that they must also play a role in describing the $SO(6)_R$ symmetry of $\mathcal{N}=4$ SYM by examining monodromy defects. Non-BPS KK monopoles in Type IIB String Theory are seven-dimensional non-BPS objects living in a space with a transverse isometry with Killing vector $\kappa$, which decay, upon tachyon condensation, into the standard BPS monopoles of Type IIB String Theory. Their existence and worldvolume action were predicted from dualities in \cite{Calvo:2025usj}. For our purposes\footnote{The action contains higher order terms which capture anomalies as in \cite{ Calvo:2025usj}. In this work, we will concentrate on the linear couplings and postpone the study of the anomalies for future work.}, in the tachyon vacuum, the action of the non-BPS KK monopole is

\begin{equation}
\label{ElectricCouplingKK}
S_{\overline{\text{KK}}_B}= \alpha\,T_{\overline{\text{KK}}_B} \int_{M_7} d(\imath_{\kappa}\,\mathcal{N}_7)\,,
\end{equation}
where $\alpha$ represents a $U(1)$ parameter associated with the integration by parts \cite{Bergman:2024aly} and $\mathcal{N}$ is related to the NSNS 6-form $\tilde{B}_6$ characterizing a type IIA background obtained by T-dualizing along the KK monopole isometry direction as 

\begin{equation}
d(\imath_\kappa \mathcal{N}_7)= e^{-2\tilde{\Phi}}\,d\tilde{B}_6\,.
\end{equation}

Following  \cite{ Calvo:2025usj}, let us first consider a KK-monopole in the coordinates of eq. \eqref{Reebcoords}. The discussion straightforwardly extends to $X_5$. In the metric of eq. \eqref{gperturbed}, we select the non-BPS KK isometry direction along $\chi$, wrapping $B$ and a three-dimensional submanifold $M_3$ in AdS$_5$. We can then read off $\tilde{B}_6$ from eq. \eqref{NSNSfielddualS5}, so that the non-BPS KK monopole action is 
\begin{equation}
S_{\overline{\text{KK}}_B}^{\text{WZ}}= -\alpha\, T_{\overline{\text{KK}}_B}\,I(B)\int_{M_3} \star dA\,,\qquad I(B)= \int_{B} g_{\chi\chi} \text{Vol}(B)\,,
\end{equation}
precisely as expected for a symmetry operator.

In the particular case of $\mathcal{N}=4$ SYM, we can translate this result into the $A^{IJ}$. Using the relation between eq. \eqref{alphacoords} and eq. \eqref{Reebcoords}, we can write

\begin{equation}
A=\frac{1}{2}K_{IJ}^{\chi}\,.
\end{equation}
The explicit expression for $K_{IJ}^{\chi}$ is rather non-illuminating. Plugging this into the generic result above, as well as the precise metric factors for this case, we finally find

\begin{equation}
\label{fiberKKmonopole}
S_{\overline{\text{KK}}}=-\alpha \,T_{\overline{\text{KK}}} \frac{\pi^2}{6}\int_{M_3}\,\star_{\text{AdS}_5} \left(dA^{12}+dA^{34}+ dA^{56}\right)\,.
\end{equation} 
Likewise, we may consider similar monopoles with different embeddings of the $\mathbb{C}P^2$ base. For instance, upon swapping
\begin{equation}
\mu^3\leftrightarrow \mu^4, \,\,\, \mu^5\leftrightarrow \mu^6\,,
\end{equation}
we would have found a non-BPS KK monopole with action

\begin{equation}
\label{fiberKKmonopole2}
S_{\overline{\text{KK}}}=-\alpha\, T_{\overline{\text{KK}}} \frac{\pi^2}{6}\int_{M_3}\,\star_{\text{AdS}_5} \left(dA^{12}-dA^{34}-dA^{56}\right)\,.
\end{equation} 
Thus, following \cite{Bah:2025vfu}, we can then construct a KK monopole measuring just $dA^{12}$ by taking the product of the non-BPS KK monopoles in eq. \eqref{fiberKKmonopole} and \eqref{fiberKKmonopole2} leading to

\begin{equation}
\label{atomiccase}
S_{\overline{\text{KK}}}=-\alpha_{\overline{\rm KK}}\, T_{\overline{\text{KK}}} \frac{\pi^2}{6}\int_{M_3}\, \star_{\text{AdS}_5}dA^{12}\,.
\end{equation}
Note that, since we are constructing a single operator through the superposition of two monopoles, it is natural to assume that each constituent contributes $\alpha=\alpha_{\overline{\rm KK}}/2$.

Given that the operator in eq. \eqref{atomiccase} measures the charge associated with the $\phi$ isometry, it is natural to ask whether one could have constructed it from a monopole with $\kappa$ along $\phi$ in \eqref{alphacoords}, wrapping $\mathcal{C}_{\phi}=\{\theta,\,S^3\}$ and the same $M_3$ in $AdS_5$. In this case

\begin{equation}
\label{dN}
d(\imath_k\mathcal{N}_7)=\frac{1}{2}\cos^3\theta\,\sin^3\theta\,K_{IJ}^{\phi}\,\left(\star_{\text{AdS}_5}dA^{IJ}\right)\wedge d\theta\wedge \omega_3\,.
\end{equation}
Hence, upon integrating over $B$, the monopole action is

\begin{equation}
\label{phiKKmonopole}
S_{\overline{\text{KK}}}=-\alpha_{\overline{\rm KK}}\, T_{\overline{\text{KK}}} \frac{\pi^2}{12}\int_{M_3}\,\star_{\text{AdS}_5}dA^{12}\,,
\end{equation}
This result is exactly $1/2$ of eq. \eqref{atomiccase}. On the other hand, the monopole with isometry along $\phi$ wraps $\mathcal{C}_{\phi}$, which is not closed. This suggests that we should regard $\mathcal{C}$ as one out of two patches making up for the actual closed space. This results in an extra factor of 2, which then makes the action equal to eq. \eqref{atomiccase}. 

\begin{figure}[h!]
\centering
\includegraphics[scale=.25]{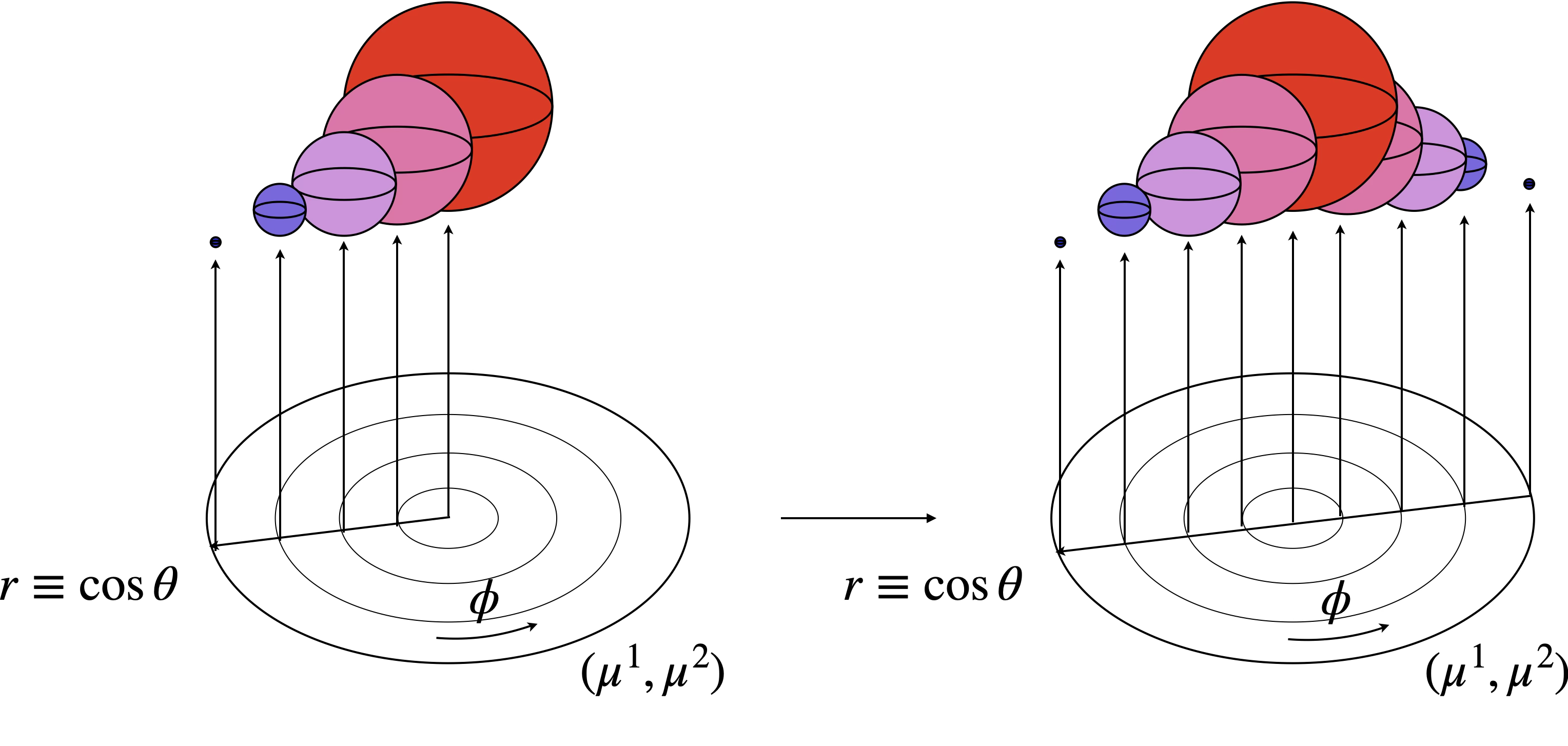}
\caption{Cartoon of the completed brane: on the left, we show the $\mathcal{C}_{\phi}$ patch. On the right, the completion to make a fully closed surface.}
\label{S5cartoon}
\end{figure}

A more formal argument is as follows. In the coordinates of eq. \eqref{alphacoords}, the $S^5$ looks like an $S^3$ fibered over a disk of radius one. The radial direction of the disk is $r=\cos\theta$, and over each $r$ there is an $S^3$ with radius $\sqrt{1-r^2}$. Thus, the $S^3$ collapses at the boundary of the disk and is maximal at its center. This disk lives in the plane $(\mu^1,\mu^2)$, and a cartoon of the $S^5$ is as in the left panel of figure \ref{S5cartoon}. In that cartoon $\mathcal{C}_{\phi}$ is literally the radial segment with the $S^3$ on top, shrinking at the boundary of the disk and staying finite at the center. It is clear that we can make a compact surface by gluing back-to-back two copies of $\mathcal{C}_{\phi}$ as on the right panel of figure $\mathcal{C}_{\phi}$. Note that at the center of the disk, the $\phi$ angle collapses. Thus, to describe in polar coordinates the full surface, we need two patches.

\subsection{4-branes}

Since the gauge fields for the R-symmetry also appear in the RR 5-form field strength, 4-branes are also natural candidates for constructing symmetry operators \cite{Bah:2025vfu}. In order to engineer a symmetry operator for $A^{12}$, we consider a non-BPS D4 brane wrapping $\mathcal{D}_{\theta\phi}=\{\theta,\phi\}$ in the coordinates of eq. \eqref{alphacoords} and a three-dimensional manifold $M_3$ on AdS$_5$. After some algebra, one finds
\begin{equation}
\label{SD4}
S_{\text{D4}}=\alpha_{\text{D4}}\,T_{\text{D4}}\, \int \frac{\cos\theta}{\sin\theta}\,\left(\star_{\text{AdS}_5}F^{12}\right)-\frac{\cos^3\theta}{\sin\theta}K^{\phi}_{IJ}\,\left(\star_{\text{AdS}_5}F^{IJ}\right)\,.
\end{equation}
Hence, upon integration over the brane worldvolume, one finally finds
\begin{equation}
S_{\text{D4}}=2\pi\,\alpha_{\text{D4}}\,T_{\text{D4}}\, \int_{M_3}\,\star_{\text{AdS}_5}dA^{12}\,.
\end{equation}
Note that, once again, $\mathcal{D}_{\theta\phi}=\{\theta,\phi\}$  is not a closed manifold. We interpret that this signals, just like for the non-BPS KK monopole, that $\mathcal{D}_{\theta\phi}$ is to be regarded as one of the patches covering a closed brane. 

Of course, one could imagine different types of 4-branes. In particular, an obviously closed brane can be constructed by considering a D4 brane wrapped on the $\mathbb{C}P^1\subset \mathbb{C}P^2$ in the coordinates of eq. \eqref{Reebcoords} and their permutations, as done in \cite{Bah:2025vfu}. It is straightforward to see that for those, one gets the same result but in terms of $\star_{\text{AdS}_5}dA^{34}$ and $\star_{\text{AdS}_5}dA^{56}$ respectively.

We note that, in addition to the non-BPS D4's that we have discussed, one could imagine all their U-dual relatives. We are led to expect that, just like the U-shaped D5 is a description of the non-BPS D4 \cite{Calvo:2025kjh,Bah:2025vfu}, the U-shaped KK-monopole in \cite{Bah:2025vfu} is the description of and object with a five-dimensional worldvolume related by duality to the D4.\footnote{Indeed, naively $T-S-T$ on a non-BPS D4 would produce an object with those properties, as can be seen by studying the effect of these transformations on the D5-$\overline{\text{D5}}$ system describing the regularization of the D4 brane.}
\section{Dielectric effect for symmetry operators}\label{dielectric}

In the previous section, we found two seemingly different operators, the non-BPS KK monopole and the non-BPS 4-brane,  measuring the same $A^{12}$ charge. It is natural to ask whether there is a relation between them. A clue can come from eq. \eqref{SD4}: even though after integration over the worldvolume the D4 only feels $\star_{\text{AdS}_5}dA^{12}$, it contains higher multipole couplings to other $A^{IJ}$ through $K_{IJ}^{\phi}F^{IJ}$. This suggests that perhaps, just like for the charged operators, the non-BPS KK-monopoles could be regarded as a dielectric expansion of an appropriate stack of 4-branes.

\subsection{Dielectric non-BPS branes}

It is known that multiple coincident non-BPS D-branes have Myers couplings which can lead to a dielectric effect for non-BPS D-branes (see \textit{e.g.} \cite{Takayanagi:2000rz}). To describe these terms for our D4 brane, let us consider firstly multiple coincident D3 branes wrapped on a four-dimensional manifold $M_4$, whose worldvolume action contains the term \cite{Myers:1999ps}
\begin{equation}
S_{\text{D3}}\supset -i\,T_{\text{D3}}\int_{M_4} {\rm Tr}\,\imath_X\imath_XC_6\,.
\end{equation}
Since D3 branes are self-dual under S-duality, it follows that there must be, as well, an analog coupling to $B_6$, the S-dual of $C_6$, as
\begin{equation}
S_{\text{D3}}\supset -i\,T_{\text{D3}}\int_{M_4} {\rm Tr}\,\imath_X\imath_XB_6\,.
\end{equation}

Let us now suppose that the branes live in a space with two isometries. One of them, denoted as $z$, is along the $M_4$ worldvolume of the brane (denoted as $M_4^z$ in the following to make explicit that it contains $z$ as an isometry). The other, denoted as $u$, is transverse to the $M_4$ worldvolume of the brane. Then, the coincident D3 brane action contains the coupling
\begin{equation}
S_{\text{D3}}\supset -i\,T_{\text{D3}}\int_{M_4^z}dx^{\mu_1}...dx^{\mu_3}dz\, {\rm Tr}\,(B_6)_{\mu_1\mu_2\mu_3ijz}\,X^iX^j \,.
\end{equation}
We can then  T-dualize along $z$, obtaining multiple coincident D2 branes with a WZ coupling

\begin{equation}
S_{\text{D2}}\supset -i\,T_{\text{D2}}\int_{M_3}dx^{\mu_1}...dx^{\mu_3} \, {\rm Tr}\,(B_6)_{\mu_1\mu_2\mu_3ijz}\,X^iX^j \,,
\end{equation}
where $M_3$ is the manifold obtained by stripping off $z$ from $M^z_4$. T-dualizing again, this time along the transverse direction $u$, gives

\begin{equation}
S_{\text{D2}}\supset -i\,T_{\text{D2}}\int_{M_4^u}dx^{\mu_1}...dx^{\mu_3}du \, {\rm Tr}\,(\mathcal{N}_7)_{\mu_1\mu_2\mu_3ijzu}\,X^iX^j \,,
\end{equation}
where $M_4^u$ is the manifold obtained by adding $u$ to $M_3$. Thus, T-duality predicts the dielectric coupling 

\begin{equation}
S_{\text{D3}}\supset -i\,T_{\text{D3}}\int_{M_4^u} {\rm Tr}\,\imath_X\imath_X \imath_z\mathcal{N}_7 \,.
\end{equation}
Hence, consistency under T-duality predicts a dielectric coupling to the KK gauge field $\mathcal{N}_7$. Since multiple coincident non-BPS D4 branes must be able to decay into multiple coincident D3 branes, we are led to conjecture the existence of an analogous coupling for the D4 brane
\begin{equation}
\label{dielectricD4}
S_{\text{D4}}\supset i\,\alpha \,T_{\text{D4}} \int_{M_5^u} {\rm Tr}\,d(\imath_X\imath_X \imath_z\mathcal{N}_7) \,,
\end{equation}
where $M_5^u$ is a five-dimensional manifold with an isometry $u$ and transverse to a space with an isometry $z$. This coupling shows that the multiple coincident D4 branes can dielectrically expand into a non-BPS KK monopole with isometric direction $z$.

\subsection{Non-BPS KK monopoles from dielectric D4's}

We now consider multiple coincident D4 branes wrapping $M_3$ in AdS$_5$ and  $\mathcal{D}_{\theta\alpha_3}=\{\theta, \alpha_3\}$. The background naturally has two isometric directions $\phi$ and $\alpha_3$. Identifying $\alpha_3$ with $u$ in eq. \eqref{dielectricD4} and $\phi$, indeed transverse to the D4's,  with $z$, from eq. \eqref{dN}, we see that
\begin{equation}
d(\imath_\phi\mathcal{N}_7)=-\cos^3\theta\,\sin^3\theta\,K_{IJ}^{\phi}\,\left(\star_{\text{AdS}_5}dA^{IJ}\right)\wedge d\theta\wedge \omega_3\,,
\end{equation}
where $\omega_3$ is the volume form of the trasverse $S^3$ spanned by $\{\alpha_1, \alpha_2, \alpha_3\}$. Switching to Cartesian coordinates $x^i$ (with $\vec{x}^2=1$) for the $S^2$ spanned by $\{\alpha_1,\alpha_2\}$, this is
\begin{equation}
d(\imath_\phi\mathcal{N}_7)=-\frac{1}{2}\sin^3\theta \cos^3\theta  K^\phi_{IJ}\,\left(\star_{\text{AdS}_5}dA^{IJ}\right)\, \wedge d\theta\wedge \epsilon_{ijk}x^idx^j \wedge dx^k\wedge d\alpha_3\,.
\end{equation}
Promoting the Cartesian coordinates $x^i$ to matrix-valued fields $X^i$, the coupling in eq. \eqref{dielectricD4} produces, after integration over $\theta$ and $\alpha_3$
\begin{equation}
S_{\text{D4}}=i\,\alpha_{\text{D4}}\, \frac{T_{\text{D4}}\pi}{3}\,\text{Tr}\left(\epsilon_{ijk}X^iX^j X^k\right)\,\int_{M_3}\,\star_{\text{AdS}_5} dA^{12}\,.
\end{equation}
Taking now the $X^i$ proportional to an $M\times M$ representation of $SU(2)$
\begin{equation}
X^i= \frac{J^i}{\sqrt{M^2-1}}\,\qquad [J^i, J^j]=2i\epsilon^{ijk} J^k\,\qquad  \vec{J}^2= (M^2-1)\,\unity\,,
\end{equation}
we finally find
\begin{equation}
S_{\text{D4}}\supset \alpha_{\text{D4}} \frac{\pi}{3}\frac{2\,M}{\sqrt{M^2-1}}T_{\text{D4}}\int_{M_3} \star_{\text{AdS}_5} dA^{12}\sim \alpha_{\rm mD4} \frac{2\pi}{3}T_{\text{D4}}\int_{M_3} \star_{\text{AdS}_5} dA^{12}\,,
\end{equation}
where in the last step we have taken the large $M$ limit and re-labelled the $\alpha$-parameter. Thus, we see that the KK monopole with isometric direction $\phi$ can be regarded as a dielectric expansion of multiple coincident D4 branes wrapping $\mathcal{D}_{\theta\alpha_3}$. Note that, just like $\mathcal{C}_{\phi}$ looked like an open manifold, $\mathcal{D}_{\theta\alpha_3}$ looks like an open manifold. As argued, the full non-BPS monopole contains two patches. Then, likewise, the dielectric expansion that we have found has to be interpreted patch-by-patch.

\section{The symTh/symTFT}\label{symTFT}

The symTh can be read off from the 5d $\mathcal{N}=8 \,\,SO(6)_R$ gauged supergravity \cite{Gunaydin:1985cu,Cvetic:2000nc}. Recall that the symTh is meant to be used only asymptotically near the AdS$_5$ boundary, where the scalars are expected to die off. Hence, the symTh can be read off from the full gauged supergravity by considering just the gauge field sector. Borrowing the gauge sector of \cite{Gunaydin:1985cu,Cvetic:2000nc}, the symTh is
\begin{equation}
\label{symThL}
\mathcal{L}_5= -\frac{1}{4\pi} \text{Tr}(\star F \wedge F)  -\frac{k}{24\pi^2} \text{Tr}\Big( A\,F^2- \frac{1}{2} A^3\,F+\frac{1}{10} A^5 \Big)\,,\qquad k=N^2\,.
\end{equation}
where $A$ is the $SO(6)_R$ R-symmetry gauge field whose (non-Abelian) field strength is $F$. We then see that the symTh of the $SO(6)_R$ symmetry can be described in terms of a 5d $SO(6)$ gauge theory with an additional Chern-Simons (CS) term. This is reminiscent of what happens for one-dimensional theories enjoying a non-Abelian symmetry, where the symTFT can be described as a free Yang-Mills theory in the bulk \cite{Witten:1992xu, Apruzzi:2024htg}. The additional Chern-Simons term has a clear interpretation as the holographic realization of the self-anomaly of the R-symmetry of 4d $\mathcal{N}=4$ SYM \cite{Witten:1998qj, Bah:2020jas}, which is present due to gaugino contributions. The coefficient is $\mathcal{A}=N^2-1\sim N^2$ at large $N$. The immediate consequence of this anomaly is the impossibility of gauging the R-symmetry or, in terms of symTh, the impossibility of imposing Neumann boundary conditions for the $SO(6)$ gauge field. At the level of symmetry operators, the presence of CS terms is in general associated in the symTFT/symTh with higher linkings among the symmetry operators describing the global symmetry. 

The charged states in eq. \eqref{symThL} are Wilson lines of $A$ in a representation $R$ of $SO(6)$. Due to Dirichlet boundary conditions, they can end at the boundary of AdS, realizing the charge operators of the dual gauge theory. For the case of $R=[0,J,0]$, these correspond to the operators described in section \ref{ChargedOps}.

Let us now discuss the symmetry operators following from eq. \eqref{symThL}. As in the end, eq. \eqref{symThL} is a standard non-Abelian gauge theory, in the bulk, we only find the familiar Gukov-Witten topological symmetry operators measuring the $SO(6)$ center.\footnote{Note that, in the presence of a CS term, the 1-form symmetry gets further reduced to $\mathbb{Z}_{\text{gcd}(k,2)}$, see \cite{Morrison:2020ool}.}
Nevertheless, in the bulk, it is still possible to construct operators measuring the charge of the Wilson lines, yet non-topological. For the case of a pure gauge theory, they were described in \cite{Cordova:2022rer} and read
\begin{equation}
\label{quasitop}
\tilde{U}_{X_0} (\Sigma_{d-1}) =\int d\mu(U) \exp \bigg \{
i \int_{\Sigma_{d-1}} {\rm Tr}
\bigg[ 
\star F U X_0 U^{-1}\bigg]\bigg \}\,,
\end{equation}
where $X_0\in \mathfrak{g}$, $U\in G$ and $\mu$ represents the Haar measure of $G$. In our case, due to the presence of the CS term, they will require additional dressing to take into account both the self-anomaly of the R-symmetry and to ensure gauge invariance, similarly to the Abelian case \cite{Calvo:2025usj}.

The symTFT offers a clearer description. Following the procedure in \cite{Mignosa:2025cpg} (which in this case can be regarded as a near-boundary limit as in \cite{Bonetti:2025dvm}), we can read off the symTFT from \eqref{symThL}
\begin{equation}
\label{symTFTSO(6)}
\mathcal{L}_5= \frac{1}{2\pi}\text{Tr}\left(B F\right)-\frac{k}{24\pi^2}\text{Tr}\left(AF^2- \frac{1}{2} A^3F+\frac{1}{10} A^5\right)\,,
\end{equation}
where $B$ is a $\mathfrak{so}(6)$ valued $d-1$ form which morally plays the role of the Hodge dual of the field strength $B\sim \star F$ in the symTFT after having performed the Hubbard-Stratonovich transformation. 

The charged operators of eq. \eqref{symTFTSO(6)} correspond to the same Wilson lines as for the symTh. On the other hand, in eq. \eqref{symTFTSO(6)} there exist topological symmetry operators in the bulk given by \cite{Bonetti:2025dvm}\footnote{Note that, on top of the symmetry operators in eq. \eqref{topopsymTFT}, we expect additional non-genuine operators attached to the AdS boundary, as described in \cite{Bonetti:2025dvm}. We are not aware of an analog to these operators in the YM theory. We leave their determination and analysis for future work.}

\begin{equation}
\label{topopsymTFT}
U_{X_0} (\Sigma_{d-1}) =\int d\mu(U) \exp \bigg \{
i \int_{\Sigma_{d-1}} {\rm Tr}
\bigg[ 
BU X_0 U^{-1}\bigg]\bigg \}\,.
\end{equation}
At the AdS boundary, due to the boundary conditions, $U_{X_0} (\Sigma_{d-1})$ further simplifies  \cite{Bah:2025vfu} and becomes

\begin{equation}
\mathcal{U}_{X_0}(\Sigma_{d-1})= e^{i \int \text{Tr}[BX_0]}\,.
\end{equation}

The symTh operators in eq. \eqref{quasitop} are the analog of the ones in eq. \eqref{topopsymTFT}. In fact, we expect that the operators in eq. \eqref{quasitop}, when pushed to the boundary, reduces to the ones in eq. \eqref{topopsymTFT} at the boundary.\footnote{More formally, in the $k\neq 0$ case, the associated dressing can be factored out in the near boundary region, leaving us with the operators in eq. \eqref{topopsymTFT}. } Thus, all in all, we expect that at the boundary, $\mathcal{U}_{X_0} (\Sigma_{d-1}) $ describes the full $SO(6)_R$ symmetry, and corresponds to the non-BPS 4-branes/non-BPS KK monopoles described above. Of course, we expect those branes to reproduce the full expression for the (quasi) topological symmetry operators, including all dressings, when pushed into the bulk. This would come from both exploiting the embedding scalars and the non-linear terms in the gauge fields, which we neglected above. We leave the determination of the full dressing and the corresponding non-linear terms for future work. 

\section{Conclusions}\label{conc}

In this note, we have studied holographic aspects of the $SO(6)_R$ R-symmetry of $4d\,\,\mathcal{N}=4$ unitary SYM. Among all operators acted by the $SO(6)_R$, we have concentrated our attention on arguably the simplest family, namely $\frac{1}{2}$ BPS scalar operators. These are in the $[0,J,0]$ representation of $SO(6)_R$, and satisfy the BPS bound $\Delta=J$. Depending on the scaling of $J$ with $N$, the holographic description is either in terms of KK fluctuations of the geometry ($J\ll N$) or in terms of giant/dual giant gravitons ($J\sim N$). To separate the discussion from the choice of metric on AdS$_5$, we have focused on giant gravitons, which are D3 branes wrapped on an $S^3\subset S^5$, spinning along a transverse $S^1\subset S^5$, and wrapping a worldline in AdS$_5$. The KK fluctuations of the geometry are naturally associated with probing GWs. This point of view allows us to bridge between the two regimes, as the D3 brane giant graviton can be regarded as the dielectric expansion of $\mathcal{O}(N)$ GW's. We then turned to symmetry operators, identifying two natural candidates. One is a non-BPS KK monopole along the lines of \cite{Calvo:2025usj}. The other is a 4-brane along the lines of \cite{Bah:2025vfu}. Insisting, for definiteness, on describing the symmetry operator associated to a particular $U(1)$ in the Cartan torus of $SO(6)_R$, say $A^{12}$, often requires to consider a superposition of branes to generate the desired worldvolume coupling to $\star_{\text{AdS}_5}dA^{12}$. One can find a more direct description where the right coupling appears straightforwardly at the price of holding only locally. With this caveat, this allows for a clear description of symmetry operators. Nevertheless, this analysis raises the question of the significance of two seemingly different operators for the same symmetry. We have argued that the resolution of this puzzle is very much parallel to the case of the charged operators. By consistency with dualities and imposing that non-BPS D4's must decay into D3 branes and reproduce their action, we have argued that the worldvolume theory for $M$ coincident non-BPS D4 branes contains the right dielectric couplings so that the D4's puff up into the non-BPS KK monopoles. Thus, the two seemingly unrelated symmetry operators are actually two sides of the same coin. In fact, this naturally suggests to link pair 4-branes with GW's and non-BPS monopoles with D3 giant gravitons. The holographic computation would really be capturing $\langle U\,O_J\rangle$, and depending on $J$, this is dominated by either a GW linking with a 4-brane ($J\ll N$) or by a D3 giant linking with a non-BPS KK monopole ($J\sim N$). It would be very interesting to shed further light on this pairing. A related context where to study analogous issues is monodromy defects, as, in field theory, they can be regarded as the endpoint of symmetry operators. At least in some cases, their holographic description seems to at least contain non-BPS KK monopoles \cite{Rodriguez-Gomez:2026mjj}. It would be very interesting to further study these cases in the light of this paper.

Having identified (a subset of) the charged operators and (at least partially) the symmetry operators for the $SO(6)_R$, it is natural to ask about the realization of the full $SO(6)$ symmetry. As argued in section \ref{symTFT}, truncating 5d $SO(6)$ gauged supergravity to the gauge sector gives a natural candidate for the symTh. Using the procedure of \cite{Bonetti:2025dvm,Mignosa:2025cpg}, we proposed a candidate for the symTFT. The operators constructed in this paper correspond to symmetry operators at the boundary. This simplifies the description, as symmetry operators at the boundary are not subject to further dressing nor additional anomaly terms due to the boundary conditions, which for the present theory are necessarily Dirichlet. 
These additional terms should arise as higher-order terms in the gauge fields of the worldvolume theory of the branes realizing the charged operators. It would be very interesting to explicitly verify the existence and explicit form of these terms and their agreement with the symTh and the symTFT operators. This could be useful in other scenarios, where the self- anomaly of the R-symmetry might allow for a different choice of boundary conditions, which would make the dressing relevant for describing the topological operators of the dual theory.

The existence of a dielectric effect for non-BPS branes can be useful to understand the fusion among topological operators and the construction of condensation defects in the context of continuous symmetries. These points were analyzed in the past in the context of BPS branes \cite{Bah:2023ymy}, which was useful to understand the analogous mechanism for discrete symmetries. Since non-BPS branes can realize continuous symmetry operators rather than discrete ones, a more precise understanding of this effect can give us useful information about the categorical aspects of continuous symmetries. Moreover, this analysis can also be useful in the discrete context, since previous evidence \cite{Bergman:2025isp} points out that non-BPS branes are also relevant to describe discrete global symmetries, at least in some contexts.

\section*{Acknowledgments}

We would like to thank Ibrahim Bah, Hugo Calvo, Mario de Marco, Michele del Zotto, Andrea Grigoletto, Jonathan Heckman, Shani Meynet, and Ethan Torres for useful discussions. F.M. also gratefully acknowledges the Uppsala TH Department, the CERN TH Department, the Brussels TH Department, and the Johns Hopkins University TH Department for hospitality while part of this work was being carried out. F.M. thanks the COST Action CA22113 "Fundamental challenges in theoretical physics" for financial support during the initial stage of this work through an STSM Grant.
D.R.-G is supported in part by the Spanish national grant MCIU-22-PID2021-123021NB-I00. 

\begin{appendix}

\section{R-symmetry from  $SO(6)$ gauged supergravity}
\label{symTh}

The $SO(6)_R$ R-symmetry of $\mathcal{N}=4$ SYM is holographically realized as the isometry group of the internal $S^5$. Therefore, the associated gauge fields in AdS$_5$ arise from gauging the $SO(6)$ isometry group. The reduction of Type IIB supergravity on the $S^5$, including all these fluctuations, was constructed in \cite{Cvetic:2000nc}. Since we are mostly interested in linear fluctuations showing the charges of probe branes, it is enough for our purposes to consider only the gauge field sector. The 10d metric including only $SO(6)_R$ gauge fields read off from \cite{Cvetic:2000nc} 

\begin{equation}
ds^2=ds_{AdS_5}^2+D\mu^I\,D\mu^I\,,\qquad D\mu^I=d\mu^I+A^{IJ}\mu^J\,,
\end{equation}
where $\mu^I$ are the embedding coordinates of $S^5$ in $\mathbb{R}^6$ satisfying $\mu^I\mu^I$. Moreover, $A^{IJ}=-A^{JI}$ are the $SO(6)_R$ gauge fields in $AdS_5$ gauging the isometries of the $S^5$. 

We stress that the gauge field fluctuations appear as well in the RR 5-form field strength. How this happens is rather cumbersome, and we refer to \cite{Cvetic:2000nc} for its explicit expression. 

\subsection{The R-symmetry gauge field}

Using the coordinates in eq. \eqref{alphacoords}, the 10d metric including the $SO(6)_R$ gauge fields is\footnote{In what follows, we are setting $g=1$ to simplify the expressions.}

\begin{align}
\label{gperturbed}
& ds^2=ds_{\text{AdS}_5}^2+ \left(d\theta- \frac{1}{2}K_{IJ}^{\theta}A^{IJ}\right)^2+\cos^2\theta\,\left(d\phi- \frac{1}{2}K_{IJ}^{\phi}A^{IJ}\right)^2+\\ \nonumber & \frac{\sin^2\theta}{4}\Bigg[ \left(d\alpha_1- \frac{1}{2}K_{IJ}^{\alpha_1}A^{IJ} \right)^2+\sin^2\alpha_1\,\left(d\alpha_2- \frac{1}{2}K_{IJ}^{\alpha_2}A^{IJ} \right)^2+\\ \nonumber
&+\left(d\alpha_3+\cos\alpha_1\,d\alpha_2-\frac{1}{2}K_{IJ}^{\alpha_3}A^{IJ}- \frac{\cos\alpha_1}{2}\,K_{IJ}^{\theta}A^{IJ}\right)^2\Bigg]\,.\nonumber
\end{align}
where $K^a_{IJ}$ are the components of the $SO(6)$ Killing vectors in terms of the angles in eq. \eqref{alphacoords} $x^a=\{\theta,\phi,\alpha_1,\alpha_2,\alpha_3\}$. They can be written using the embedding coordinates as

\begin{equation}
K_{IJ}^{a}=g^{ab}\,(\mu^I\partial_b\mu^J-\mu^J\partial_b\mu^I)\,,
\end{equation}
with $a$ indices raised and lowered with the unperturbed metric.\footnote{Embedding space indices $I,J$ are raised and lowered with $\delta$, so there is really no distinction between upper and lower indices.} The explicit expression in terms of the angles is rather cumbersome. For instance
\begin{align}
\label{Kphi}
&\frac{1}{2}K^{\phi}_{IJ}A^{IJ}=A^{12}+\\ 
&\nonumber  +A^{16} \sin \left(\frac{\alpha_1}{2}\right) \tan \theta \sin \phi \sin \left(\frac{\alpha_3-\alpha_2}{2}\right)-A^{23}\cos \left(\frac{\alpha_1}{2}\right) \tan \theta \cos \phi \cos \left(\frac{\alpha_2+\alpha_3}{2}\right)\\ 
&\nonumber +A^{13} \cos \left(\frac{\alpha_1}{2}\right) \tan \theta \sin \phi \cos \left(\frac{\alpha_2+\alpha_3}{2}\right)+A^{14} \cos \left(\frac{\alpha_1}{2}\right) \tan \theta \sin \phi \sin \left(\frac{\alpha_2+\alpha_3}{2}\right)\\ 
&\nonumber +A^{15} \sin \left(\frac{\alpha_1}{2}\right) \tan \theta \sin \phi \cos \left(\frac{\alpha_3-\alpha_2}{2}\right)-A^{24} \cos \left(\frac{\alpha_1}{2}\right) \tan \theta \cos \phi \sin \left(\frac{\alpha_2+\alpha_3}{2}\right)\\
&\nonumber -A^{25} \sin \left(\frac{\alpha_1}{2}\right) \tan \theta \cos \phi \cos \left(\frac{\alpha_3-\alpha_2}{2}\right)+A^{26} \sin \left(\frac{\alpha_1}{2}\right) \tan \theta \cos \phi \sin \left(\frac{\alpha_2-\alpha_3}{2}\right)\,.
\end{align}

One can check that $K_{IJ}=K_{IJ}^{a}\partial_a$ satisfy the $SO(6)_R$ algebra

\begin{equation}
[K_{AB},K_{CD}]=\delta_{BC}\,K_{AD}-\delta_{AC}\,K_{BD}-\delta_{BD}\,K_{AC}+\delta_{AD}\,K_{BC}\,.
\end{equation}

A similar expansion can be done in the coordinates in eq. \eqref{Reebcoords}. 
Instead of writing the generic expression, let us only consider gauging the Reeb isometry and look at the resulting metric, which extends to generic $X_5$. Then, the metric including the gauge field fluctuation is
\begin{equation}
\label{gperturbedReeb}
ds^2=ds_{\text{AdS}_5}^2+g_{\chi\chi}\left(d\chi+\xi+ \frac{2}{3}\,g_{\chi\chi}^{-\frac{1}{2}}\,A\right)^2+ds_B^2\,.
\end{equation}

\end{appendix}

\bibliography{ArXiv_v1bib}
\bibliographystyle{JHEP}

\end{document}